\documentclass[onecolumn, amsmath, amssymb, superscriptaddress]{revtex4}

\usepackage{dcolumn}
\usepackage{bm}
\usepackage{amsmath}
\usepackage{amsfonts}
\usepackage{mathrsfs}  
\usepackage[breaklinks=true, hyperfootnotes=false]{hyperref}
\usepackage{graphics}
\usepackage[dvips]{graphicx}
\usepackage{times}
\usepackage{setspace}
\usepackage{color}   
\usepackage{verbatim}
\usepackage[raggedright]{titlesec}
\usepackage[normalem]{ulem}
\usepackage{framed}
\usepackage{enumerate}
\usepackage{breakurl}
\usepackage{float}

\usepackage[graphicx]{realboxes}
\usepackage{varwidth} 
\usepackage{balance}
\usepackage{flushend}

\hypersetup{colorlinks=true,citecolor=blue, linkcolor=black,urlcolor=blue} 
\usepackage[outercaption]{sidecap} 

\usepackage{hyperref}
\usepackage{amssymb}
\usepackage{mathbbol}
\usepackage{flushend}

\renewcommand{\thesection}{\arabic{section}}

\newcommand{\beginsupplement}{%
        \setcounter{section}{0} 
        \renewcommand{\thesection}{A\arabic{section}}%
        \setcounter{table}{0}
        \renewcommand{\thetable}{A\arabic{table}}%
        \setcounter{figure}{0}
        \renewcommand{\thefigure}{A\arabic{figure}}%
        \setcounter{equation}{0}
        \renewcommand{\theequation}{A\arabic{equation}}%
     }

\begin{document}
\title{The disruption index  is biased by citation inflation} 
\author{Alexander Michael Petersen$^{a,}$}
\author{Felber Arroyave}
\affiliation{Department of Management of Complex Systems, Ernest and Julio Gallo Management Program, School of Engineering, University of California, Merced, California 95343, USA}
\author{Fabio Pammolli}
\affiliation{Politecnico Milano, Department of Management, Economics and Industrial Engineering, Via Lambruschini, 4/B, 20156, Milan, Italy}


\maketitle

\footnotetext[1]{ \ \ Send correspondence to:  apetersen3@ucmerced.edu}

\vspace{-0.2in}
{\bf \noindent A recent analysis of scientific publication and patent  citation networks by Park et al. (Nature, 2023)  suggests that publications and patents are becoming less disruptive over time.  
Here we show that the reported  decrease in disruptiveness  is  an artifact of  systematic  shifts in the structure of citation networks unrelated to  innovation system capacity. 
Instead, the decline is attributable to  `citation inflation', an unavoidable characteristic of real citation networks that manifests as a systematic time-dependent bias and  renders cross-temporal analysis  challenging. 
One  driver of citation inflation is the ever-increasing lengths of reference lists over time, which in turn increases the density of links in  citation networks, and causes the disruption index to converge to 0.
A second driver is attributable to  shifts in the construction of reference lists, which is increasingly impacted by self-citations that increase in the rate of triadic closure in citation networks, and thus   confounds efforts to measure disruption, which is itself a measure of triadic closure. 
Combined, these two systematic shifts render  the disruption index temporally biased, and unsuitable for cross-temporal analysis.  
The impact of this systematic bias further stymies efforts to correlate disruption to other measures that are also time-dependent, such as team size and citation counts. 
In order to demonstrate this fundamental measurement problem, we  present three complementary lines of critique (deductive, empirical and computational modeling), and also make available an ensemble of synthetic citation networks that can be used  to test alternative  citation-based indices for  systematic bias.}\\

A measure of disruption  was recently developed and applied to empirical citation networks \cite{funk2017dynamic,wu2019large,park2023papers}. 
This bibliometric measure, denoted by $CD$, quantifies the degree to which an intellectual contribution $p$ (e.g. an research publication or invention patent)  supersedes the sources cited in its reference list, denoted  by  the set $\{r\}_{p}$. 
As defined, $CD_{p}$ is measured according to the local  structure of the subgraph $G_{p}=\{r\}_{p}\cup p \cup  \{c\}_{p}$ comprised of the focal node $p$, nodes belonging to its reference list $\{r\}_{p}$, and  the set of nodes citing either $p$ or any member of $\{r\}_{p}$, denoted by   $\{c\}_{p}$. 
If future  intellectual contributions cite $p$ but do not cite members of $\{r\}_{p}$, then it is argued that $p$ plays a disruptive role in the citation network. 
However, the critical issue we  highlight is the following: as the length $r_{p} = \vert  \{r\}_{p} \vert$ of the reference list  increases, so does the likelihood that one of those papers is highly cited. Hence,  $CD_{p}$ is a biased measure because  reference lists have increased dramatically over time, and so too have the number of citations that highly-cited papers accrue \cite{pan2016memory} -- both phenomena being  bi-products of   {\it citation inflation} \cite{petersen_citationinflation_2018}.  

Citation inflation (CI) refers to the systematic increase in the number of links introduced to the scientific (or patent) citation network each year. CI is analog to monetary inflation \cite{orphanides1990money,orphanides2003quest}, whereby as a government prints more money the sticker price of items tends to go up, rendering the  impression that the real cost of goods are increasing (to what degree this relationship is valid depends on  wage growth and a number of other economic factors). 
By analogy, it might also be tempting to attribute the increased volume of scientific production to techno-social productivity increases, yet this explanation neglects the persistent growth rate of the inputs (e.g. researchers and research investment) that are fundamental to the  downstream production of outputs  (e.g.  research articles, patents.

Indeed, secular growth underlies various quantities relevant to the study of  the scientific endeavor, from national expenditures in  R\&D to the population size of researchers \cite{petersen_citationinflation_2018} and the characteristic number of authors  per research publication \cite{Wuchty:2007,petersen_quantitative_2014,pavlidis_together_2014} -- all quantities that have persistently grown over the last century.
Nevertheless, the degree to which such  growth  affects the quantitative evaluation  of research outcomes is under-appreciated,   
and can manifest in inconsistent measurement frameworks and metrics.  
Indeed, the number of  citations an article receives is not solely attributable to novelty or prominence of the research, but
also depends on the the population size and citing norms of a discipline, and quite fundamentally,  the nominal production rate of links in the citation network, among other considerations \cite{bornmann2008citation}.
 Hence, there is real need to distinguish nominal counts versus real values in scientific evaluation, which in the analysis of citation networks requires accounting for   {\it when} each citation was produced, and in further extensions, {\it how}  the credit is  shared \cite{petersen2010methods,pavlidis_together_2014,petersen_citationinflation_2018,shen2014collective}. 
 
\begin{SCfigure}
\includegraphics[width=0.52\textwidth]{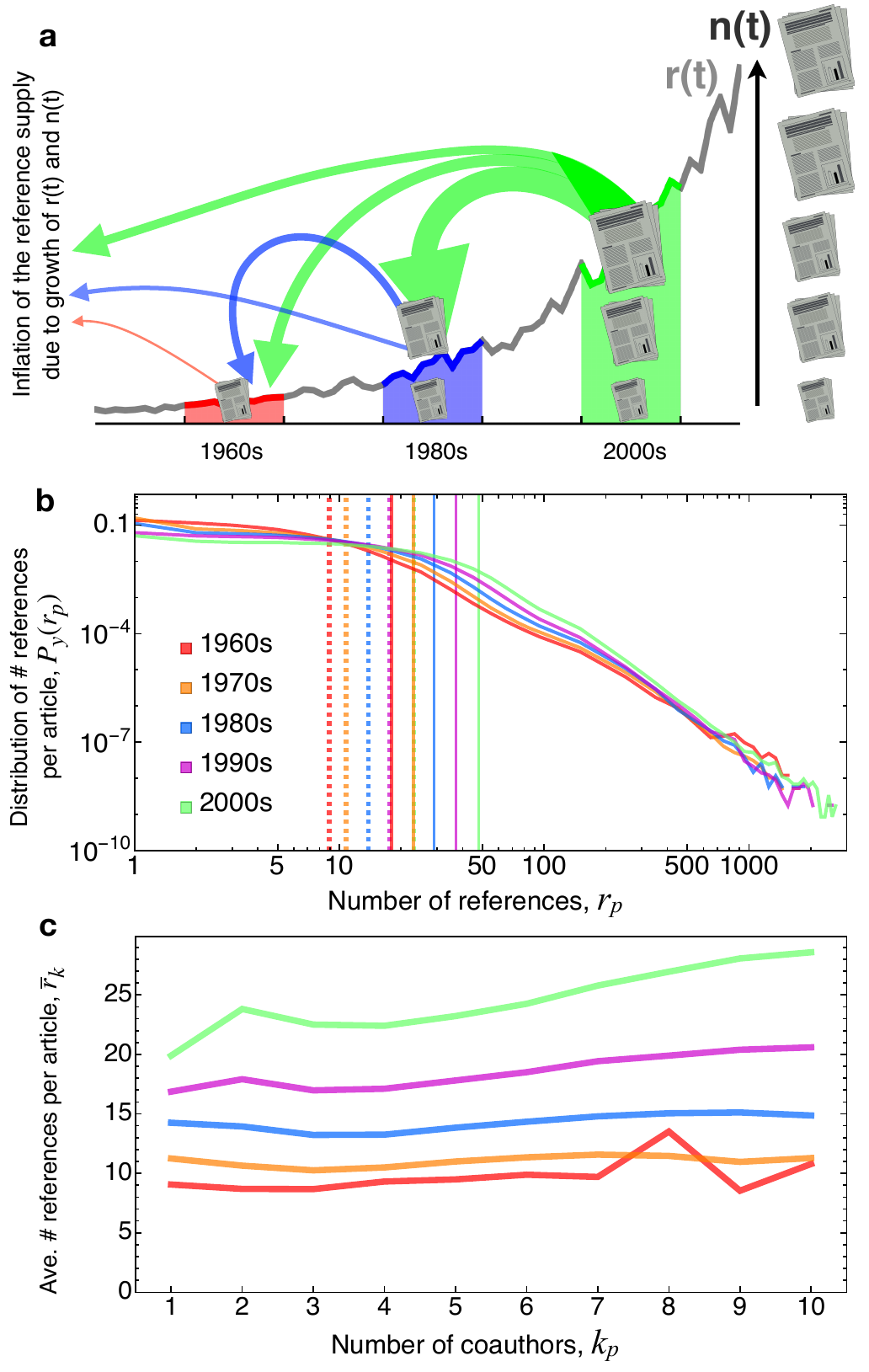}
 \caption{  \label{Figure1.fig} {\bf  `Citation inflation' attributable to the increasing number and length of reference lists.}
  (a) Schematic illustrating the inflation of the reference supply  owing to the fact that the annual publication rate $n(t)$ (comprised of increasing diversity of article lengths), along with the number of references per publication $r(t)$, have grown exponentially over time $t$, which implies a non-stationary cross-generational flow of attribution in real citation networks. Such citation inflation cannot be controlled  by way of fixed citation windows \cite{petersen_citationinflation_2018}.
 (b) The probability density function $P_{y}(r_{p})$ of the number of references per article $r_{p}$ calculated for articles included in the MAG citation network grouped by the decade of publication $y$. Vertical dashed lines indicate the average value; vertical solid lines indicate the 90th percentile, such that only the 10\% largest $r_{p}$ values are in excess of this value.
 (c) Conditional relationship between two quantities that systematically grow over time ($k_{p}$ and $r_{p}$). Note the increasing levels and slope of the relationship over the 50-year period. This relationship indicates that   regressing $CD_{p}$ on $k_{p}$ -- while omitting $r_{p}$ as a covariate and thereby neglecting the negative relationship between $CD_{p}$ and $r_{p}$ --   may lead to the  confounded conclusion that $CD_{p}$ decreases as $k_{p}$ increases. 
 }
\end{SCfigure}

So what are the main sources of CI in scientific citation networks and what are the real-world magnitudes of their effects?  {\bf Figure \ref{Figure1.fig}(a)} illustrates how CI  arises through the combination of longer reference lists, denoted by $r(t)$,   compounded by an increasing production volume, $n(t)$. 
By way of real-world example, prior calculation of the growth rate of total number of citations produced per year based upon the entire Clarivate Analytics Web of Science citation network estimated that the total volume $C(t) \approx n(t)r(t)$ of  citations  generated by the scientific literature grows exponentially with annual rate   $g_{C} = g_{n} + g_{r}  = 0.033 + 0.018 = 0.051$ \cite{pan2016memory}. 
Hence, with the number of links in the citation network growing by roughly 5\% annually, the total number of links in the citation doubles every  $\ln(2)/g_{C} = 13.6$ years!

While the dominant contributor to CI is the growth of $n(t)$ deriving from increased researchers and investment in science coupled with technological advances increasing the rate of manuscript production, the shift away from print towards online-only journals, and the advent of multidisciplinary megajournals \cite{petersen2018mismanagement},  the contribution to CI from growing reference lists alone is nevertheless  substantial and varies by discipline \cite{sanchez2018reference,abt2002relationship,dai2021literary,nicolaisen2021number}. 
 By way of example, consider descriptive statistics based upon analysis of millions of research publications comprising the Microsoft Academic Graph (MAG) citation network  \cite{sinha2015overview}: in the 1960s,  the average ($\pm$ standard deviation) number of references per articles was $\overline{r_{p}}=9$ ($\pm$17); by the 2000s, $\overline{r_{p}}$ increased to 23 ($\pm 27$), a 2.6-fold increase over the 50-year period  -- see {\bf Fig. \ref{Figure1.fig}(b)}. 
 Meanwhile, as research team sizes -- denoted by $k_{p}$, and used as a proxy for the production effort associated with a research output -- increase in order to address research problems featuring greater topical and methodological breadth, there emerges a non-linear relationship between $\overline{r}_{p}$ and $k_{p}$ showing that the modern research article  is fundamentally different from those produced even a decade ago --  see {\bf Fig. \ref{Figure1.fig}(c)}. 
  Thus, not only does the nominal value of a citation vary widely by era, but the implications of secular growth on the topology of the citation network and thus citation-based research evaluation are profound \cite{pan2016memory}. 
  A standard solution to taming variables that are susceptible to  inflation is to use a deflator index, which amounts to normalizing the cross-temporal variation by way of standardized reference point \cite{petersen2010methods,petersen_citationinflation_2018}. 
 Another more nefarious problem is the accurate measurement of the quantitative relationship between variables that are independently   growing over time, which is susceptible to  omitted variable bias if the role of time is neglected.
  
In what follows, we demonstrate the implications of CI that render  $CD$  unsuitable for cross-temporal analysis, and call into question the empirical analysis and  interpretations of trends in scientific and technological advancement based upon $CD$ \cite{wu2019large,park2023papers}. To establish how the disruption index suffers from citation inflation and is confounded by shifts in scholarly citation practice, we employ three different approaches: deductive analysis based upon the definition of $CD_{p}$, empirical analysis of the Microsoft Academic Graph (MAG) citation network, and computational modeling of synthetic citation networks. In the latter approach,  we are able to fully control the sources of the systematic bias underlying $CD$ (namely CI), thereby demonstrating that  $CD$ follows a stable frequency distribution in the absence of CI. We conclude with research evaluation policy implications.

\vspace{-0.2in}
\section{Quantitative definition of $CD$ and a deductive critique}
\begin{figure*}
\centering{\includegraphics[width=0.85\textwidth]{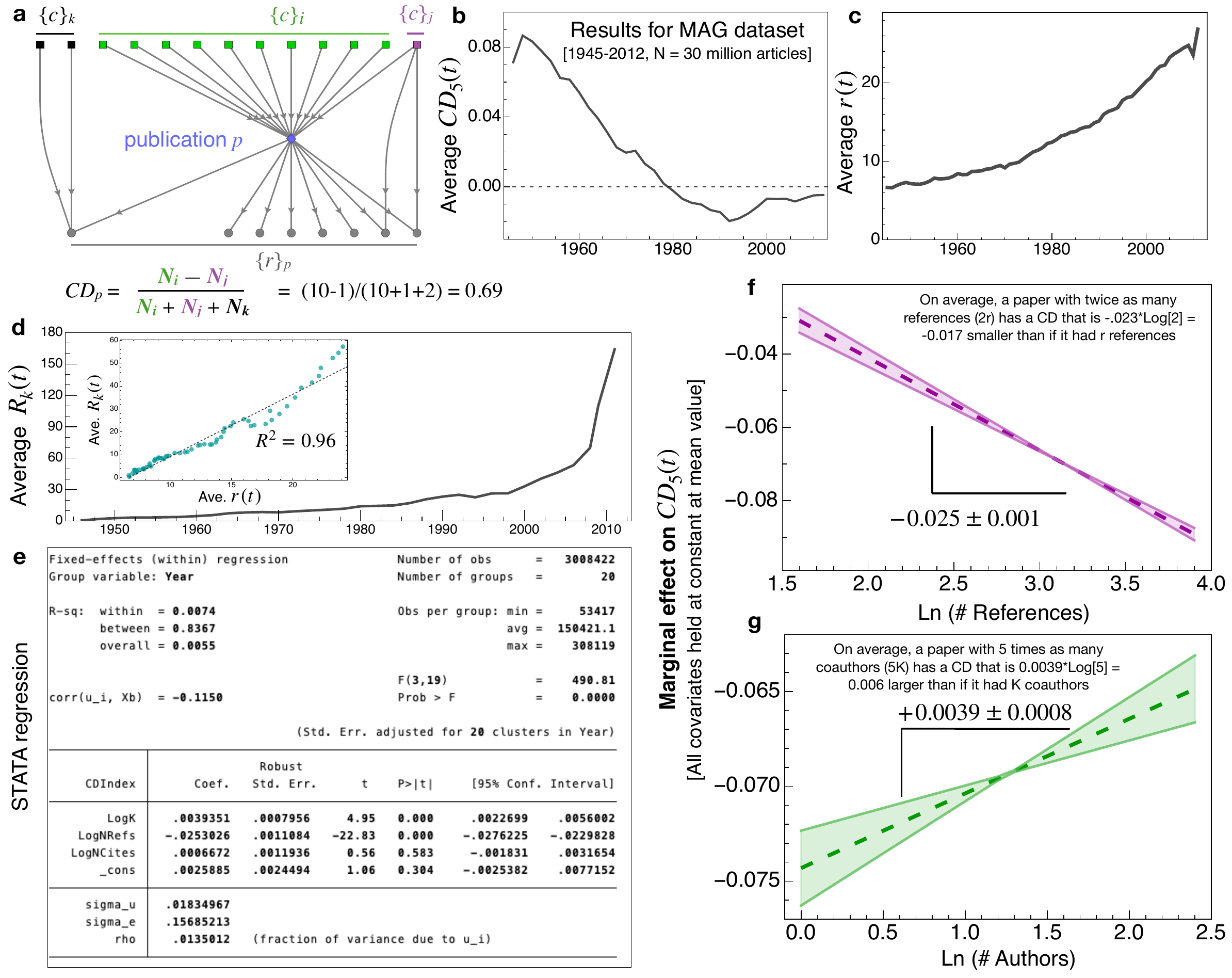}}
 \caption{  \label{Figure2.fig} {\bf Empirical analysis of the disruption index.}
 (a)  Schematic of the disruption index calculation based upon the sub-network revolving around the source publication/patent $p$.  The disruption index $CD_{p}$ can  be calculated by identifying three non-overlapping subsets of $\{c\}_{p} = \{c\}_{i} \cup \{c\}_{j} \cup \{c\}_{k}$, of sizes $N_{i}$, $N_{j}$ and $N_{k}$, respectively. The  subset $i$ refers to members of $\{c\}_{p}$ that cite the focal $p$ but do not cite any elements of $\{r\}_{p}$, and thus measures the degree to which $p$  disrupts the flow of attribution to foundational members of $\{r\}_{p}$. The  subset $j$ refers to  members of $\{c\}_{p}$ that cite both $p$ and $\{r\}_{p}$, measuring the degree of consolidation that manifests as triadic closure in the subnetwork (i.e., network triangles  formed between $p$,  $\{r\}_{p}$, $\{c\}_{j}$). The subset $k$ refers to members of $\{c\}_{p}$ that cite $\{r\}_{p}$ but do not cite $p$.  
 (b) Average  disruption index, $CD_{5}(t)$  calculated  using a 5-year citation window based upon $29.5\times 10^{6}$ articles from the MAG dataset from 1945-2012. 
 (c) Average number of references per paper per year, $r(t)$, which increased by a factor of 4 over the 6-year period shown. 
 (d) Average extraneous citation rate, $R_{k}(t) \gg 1$ that is central to the critique of $CD$, and derives from the increasing citation count of highly-cited papers belonging to the reference list $\{r\}_{p}$ which systematically inflates the size of the extraneous set $\{c\}_{k}$.
  (inset)  $R_{k}(t)$ grows roughly proportional to $r(t)$.
 (e) Results of linear regression model implemented in STATA 13 for dependent variable $CD_{p,5}$,  controlling for $r_{p}$ and secular growth by way of yearly fixed-effects. Publication years are within the 20-year range 1990-2009; covariates are included following a logarithmic transform. 
   (d) Marginal effects calculated with all other covariates  held at their mean values, showing that $CD_{5}$ is negatively correlated with the log of the number of references, $\ln r_{p}$. (e) $CD_{5}$ is positively correlated with   the log of the number of coauthors, $\ln k_{p}$.   }
\end{figure*}

\vspace{-0.2in}
The disruption index is a higher-order network metric that incorporates information extending  beyond the first-order links connecting to $p$ -- those nodes that cite $p$ and are prospective (forward looking or diachronous), and those nodes that are referenced by $p$, and thus retrospective (backward looking or synchronous)    \cite{Nakamoto_synchronous_1988,Glanzel_forwardbackward_2004}.
The original definition of $CD$ was formulated as a conditional sum across the adjacency matrix \cite{funk2017dynamic}, and was subsequently reformulated as a ratio \cite{wu2019large}. According to the latter conceptualization, calculating $CD_{p}$  involves first identifying three non-overlapping subsets of citing nodes, $\{c\}_{p} = \{c\}_{i} \cup \{c\}_{j} \cup \{c\}_{k} $, of sizes $N_{i}$, $N_{j}$ and $N_{k}$, respectively -- see {\bf Fig. \ref{Figure2.fig}(a)}  for a schematic illustration. 

The  subset $i$ refers to members of $\{c\}_{p}$ that cite the focal $p$ but do not cite any elements of $\{r\}_{p}$, and thus measures the degree to which $p$  disrupts the flow of attribution to foundational members of $\{r\}_{p}$. The  subset $j$ refers to  members of $\{c\}_{p}$ that cite both $p$ and $\{r\}_{p}$, measuring the degree of consolidation that manifests as triadic closure in the subnetwork (i.e., network triangles  formed between $p$,  $\{r\}_{p}$, $\{c\}_{j}$). The subset $k$ refers to members of $\{c\}_{p}$ that cite $\{r\}_{p}$ but do not cite $p$. 
As such, the $CD$ index is given by the  ratio,
\begin{eqnarray}
CD_{p} = \frac{N_{i}-N_{j}}{N_{i}+N_{j}+N_{k}} ,
\end{eqnarray}
which can be rearranged as follows,
\begin{eqnarray}
CD_{p} = \frac{(N_{i}-N_{j})/(N_{i}+N_{j})}{1+N_{k}/(N_{i}+N_{j})}  = \frac{CD_{p}^{\text{nok}}}{1+R_{k}} \ .
\end{eqnarray}
The ratio  $R_{k} = N_{k}/(N_{i}+N_{j}) \in [0, \infty)$ is an extensive quantity that measures the rate of extraneous citation, whereas  $CD_{p}^{\text{nok}} \in [-1,1]$ is an intensive quantity.  The polarization measure $CD_{p}^{\text{nok}}$ is an alternative definition of disruption that simply neglects $N_{k}$ in the denominator  \cite{bornmann2020disruption}; for this reason, characteristic values of $CD_{p}^{\text{nok}}(t)$ are larger and decay more slowly over time then  respective $CD_{p}(t)$ values -- see ref. \cite{park2023papers}. 
Following initial criticism regarding the definition of $CD_{p}$ \cite{wu2019confusing,bornmann2020disruption},  other variations on the theme of $CD$ have since been  analyzed  \cite{wu2019large} and critiqued according to their advantages and disadvantages \cite{leydesdorff2021proposal}. 

To summarize, we argue that a  simple deductive explanation trumps the  alternative socio-technical explanations offered \cite{park2023papers,kozlov2023disruptive} for the decline in $CD$ calculated for  publications and patents.  
Namely, the disruption index $CD_{p}$ systematically declines, along with similar $CD_{p}$ variants \cite{bornmann2020disruption,leydesdorff2021proposal,wu2019large}, for the simple reason that $CD$ features a numerator that is bounded and a denominator that is unbounded.  More technically, the term $R_{k}$ is susceptible to CI,  which is entirely sufficient to explain why  $CD$  converges to 0 over time. 

\vspace{-0.2in}
\section{Empirical critique}
\vspace{-0.2in}
In this section we show empirically  that  $CD_{p}$  declines over time due to the  runaway growth of  $R_{k}(t)$, and implicitly, $r(t)$.
While our results are based upon a single representation of the scientific citation network made openly available by the MAG project  \cite{sinha2015overview}, the implications are generalizable to citation networks featuring CI characterized by a non-stationary number of new links introduced by each new cohort of new citing items.
 To be specific, the citation network we analyzed is formed from the roughly $29.5\times 10^{6}$ million research articles in the MAG  dataset that have a digital object identifier (DOI), were published between 1945-2012, and belonging to a  mixture of research areas.
 
{\bf Figure \ref{Figure2.fig}}(a) shows a schematic of the  sub-graph used to calculate the $CD_{p}$ value for each publication. To be consistent with \cite{wu2019large,park2023papers}, we calculate $CD_{p,CW}(t)$  using a $CW=5$-year citation window (CW), meaning that only articles published within 5 years of $p$ are included in the subgraph $\{c\}_{p} =\{c\}_{i} \cup \{c\}_{j} \cup \{c\}_{k}$.  
 As such, {\bf Fig. \ref{Figure2.fig}}(b) shows a decline in the average  $CD_{5}(t)$  that is consistent with the overall trend shown in Fig. 2    in ref. \cite{park2023papers}, where the data are disaggregated by discipline; also note that   Fig. 2 and  ED Figs. 6 and 9   in \cite{park2023papers} show that disciplines with higher publication volumes and thus more references produced (life sciences and biomedicine, and physical sciences) tend to have smaller $CD_{5}(t)$ values in any given year relative to the social sciences (e.g. JSTOR), which is qualitatively consistent with our critique.  

 We also note that while  the implementation of a CW may control for right-censoring bias, it does not  control  CI in any precise way. By way of example, consider the impact of the CW on  $N_{k}$, the number of extraneous articles that do not cite $p$ but do cite elements of $\{r\}_{p}$.
 A CW will reduce the number of papers contributing to $CD_{5}(t)$ via $N_{k}$, but it will also reduce  $N_{i}+N_{j}$ in similar proportions, leaving the ratio $R_{k}(t)$ unchanged, on average. 
 Consider a more quantitative explanation that starts by positing that $N_{k}$ increases proportional to $n(t)r(t)$, as the nodes belonging to $\{c\}_{k}$ are  unconstrained by the first-order citation network  $\{c\}_{i} \cup \{c\}_{j}\cup \{r\}_{p}$. Following the same logic, $N_{i}+N_{j}$ grows proportional to $n(t)$. In both cases, even if the proportionality constant depends weakly on $CW$, the ratio $R_{k}(t)$  will grow proportional to $r(t)$.
  
  There is likely to be considerable variance in the publication-level relationship  between $R_{k,p}$ and $r_{p}$, because if any member of $\{r\}_{p}$ is highly cited, then $N_{k}$ is skewed towards the heavy right tail of the citation distribution. Moreover, the base number of citations associated with extreme values in the citation distribution have increased dramatically over the last half century as a result of CI, such that the number of citations $C(Q \vert t)$ corresponding to the $Q=$99th percentile of the citation distribution  increased at an annual rate of roughly 2\% from roughly 55 citations in 1965 to roughly 125 citations in 2005 -- see Fig. 4 in ref. \cite{pan2016memory}. For this reason, the term $N_{k}$ introduces  susceptibility to CI according to two channels.

Here we focus on the channel associated with the growth of $r(t)$,  which grew at roughly the same rate as $C(99 \vert t)$, growing from roughly 9 to 23 references per paper over the same period --  see {\bf Fig. \ref{Figure2.fig}}(c).
 Consequently,  $R_{k}(t) \gg 1$ for nearly the entire period of analysis and that the growth of $R_{k}(t)$ is largely explained by the growth of $r(t)$ in the empirical data -- see  {\bf Fig. \ref{Figure2.fig}}(d). For this reason, it is  more accurate to describe $CD$ as converging to 0 as opposed to decreasing over time. 

In order to confirm  these aggregate-level relationships at the publication level, we applied a linear regression model whereby the unit of analysis is an individual publication.
The  linear  model specification is given by
\begin{eqnarray}
CD_{p,5} = b_{0} + b_k \ln k_{p} + b_r \ln r_{p} + b_c \ln c_p   + D_t + \epsilon_t
\end{eqnarray}
which controls for secular growth by way of yearly fixed-effects, denoted by $D_{t}$. 
The results of the ordinary least squares (OLS) estimation using the STATA 13.0 package xtreg are shown in {\bf Fig. \ref{Figure2.fig}}(e), and are based upon 3 million publications with $1 \leq k_{p} \leq 10$ coauthors, $5 \leq r_{p} \leq 50$ references, and $10 \leq c_{p} \leq 1000$ citations that were published in the two-decade period 1990-2009. 
The independent variables are modeled using a logarithmic transform because they are each  right-skewed: ``LogK'' corresponds to $ \ln k_{p}$; ``LogNRefs'' corresponds to $\ln r_{p}$; and LogNCites  corresponds to $\ln c_{p}= \ln(c_{i}+c_{j})$,  the number of citations received by $p$ in the 5-year window. 
This sample of MAG articles were used so that results are more closely comparable to  Wu et al. who focus on articles with $k_p \in [1,10]$ \cite{wu2019large}.

Results indicate a negative relationship between $CD_{p,5}$ and the number of references, consistent with our deductive argument. {\bf Figure \ref{Figure2.fig}}(f) shows the marginal relationship with $\ln r_{p}$, holding all covariates at their mean values, and indicates a net shift in $CD$ of roughly -0.06 units  as $r_{p}$ increases by a factor of 10 from 5 to 50 total references. 
Similarly, {\bf Fig. \ref{Figure2.fig}}(g) shows the marginal relationship with $\ln k_{p}$, indicating a net shift in $CD$ of roughly +0.01 units  as $k_{p}$ increases by a factor of 10 from 1 to 10 coauthors, which is in stark contrast to the  relationship with opposite sign reported in ref. \cite{wu2019large}.

\vspace{-0.2in}
\section{Computational critique}
\vspace{-0.2in}
\subsection{Generative network model featuring citation inflation and redirection}
\vspace{ -15 pt}
We employ computational modeling to  explicitly control several fundamental sources of variation, and to also explore complementary mechanisms contributing to shifts in $CD$ over time --  namely,  shifts in scholarly citation practice.
Our identification strategy is to  growth synthetic citation networks that are identical in growth trajectory and size, but differ just in the specification of (i) $r(t)$ and/or (ii) the rate of triadic closure denoted by $\beta$ that controls the consolidation-disruption difference  defining the numerator of $CD$. 

We model the  growth of a citation network using  a model originally developed in ref. \cite{pan2016memory} that applies  Monte Carlo (MC) simulation to operationalize  stochastic link  dynamics by way of a random number generator.  
This model  belongs to the class of growth and redirection models \cite{krapivsky_network_2005,barabasi2016network}, and reproduces a number of statistical regularities established for real citation networks -- both structural (e.g. a log-normal citation distribution \cite{UnivCite}) and dynamical (e.g., increasing reference age with time \cite{pan2016memory}; exponential citation life-cycle decay \cite{petersen_reputation_2014}).  -- see the {\it Appendix Section A1} and {\bf Fig. \ref{FigureS1.fig}} for more information regarding the empirical validation of our generative network model. 
The synthetic networks constructed and analyzed in what follows  are openly available \cite{DryadDisruption2023} and can be used to test $CD$ and  other citation-network based bibliometric measures for sensitivity to CI and other aspects of secular growth.
 
We construct each synthetic citation network by sequentially adding new layers of nodes of prescribed volume  $n(t)$  in each MC period  $t\geq0$ representing a publication year. 
Each new node, denoted by the index $a$,  represents a publication that could in principal cite any of the other existing nodes in the network.
As such, the resulting synthetic networks are representative of a single scientific community, and also lack latent node-level variables identifying disciplines, authors, journals, topical breadth or depth, etc. 

\begin{figure*}
\centering{\includegraphics[width=0.99\textwidth]{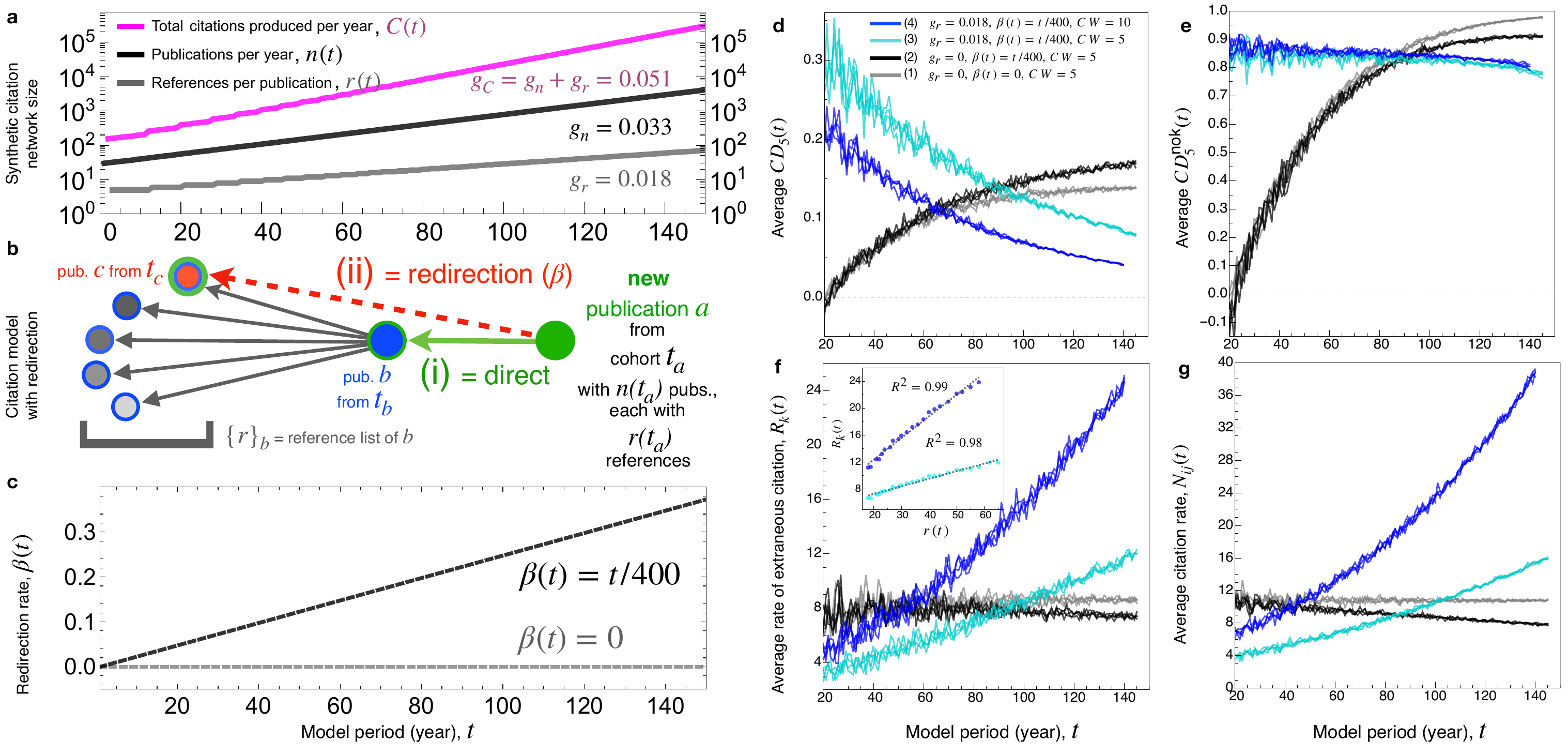}}
 \caption{  \label{Figure3.fig}  {\bf Numerical  simulation of growing citation networks elucidates roles of citation inflation and strategic citation practice.}
 (a) Model system evolved over $T=150$  periods (representing years), using growth parameters estimated for the entire Clarivate Analytics  Web of Science citation network \cite{pan2016memory}. 
  (b) Schematic of the citation model comprised of two citing mechanisms: (i) direct citations, and (ii) redirected citations made via the reference list  $\{r\}_{b}$ of an intermediate item $b$. Type (ii) references give rise to triadic closure corresponding to the $N_{j}$ factor in $CD_{p}$. 
  (c) The rate of type (ii) references is controlled by the  parameter $\beta(t)$, which quantifies the fraction of links in the citation network directly following this  `consolidation' mechanism  \cite{funk2017dynamic,park2023papers}, which yields more negative $CD_{p}$ values. To disentangle the roles of citation inflation (owing to $g_{r}>0$) from shifts in scholarly citation practice (owing to  $\partial_{t}\beta(t)>0$), we compare four scenarios: scenarios (1,2) (gray and black curves) feature no citation inflation $(g_{r}=0)$; (2,3)  compare  $\beta(t)=0$ and $\beta(t) = t/400$; and (3,4) (cyan and blue curves) compare the effects of different citation windows (CW). 
   (d) Each curve is the average $CD_{CW}(t)$ calculated for a single synthetic network. (e)  Average $CD^{\text{nok}}_{5}(t)$. 
 (f) $R_{k}(t)$ is the average rate of extraneous citations, which  increases as either $r(t)$ or CW  increase. (inset) High linear correlation between $r(t)$ and $R_{k}(t)$ shows that the decreasing trend in $CD(t)$ is largely attributable to citation inflation. 
 (g) The average value of  $N_{ij}(t)  = N_{i} +N_{j}$ (which  defines the denominator of  $CD^{\text{nok}}$)  also systematically increases, and so neglecting the term $N_{k}$ does not solve the fundamental issue of CI. }
\end{figure*}

We seed the network with  $n(t=0) \equiv 30$  `primordial' nodes that are disconnected, i.e. they have reference lists of size $r_{a} \equiv 0$. This ensures that the initial conditions are the same for all networks generated. All nodes added thereafter have reference lists of a common prescribed size, denoted by $r(t)$. These rules ensure there is no variation within a given publication cohort  regarding their synchronous connectivity.
To model the exponential growth of scientific production, we  prescribe the number of new ``publications'' according to the exponential trend $n(t)=n(0)\exp[g_{n}t]$. We use $g_{n} \equiv 0.033$ as the  publication  growth rate  empirically derived  in prior work \cite{pan2016memory}.
 Similarly, we  prescribe the number $r(t)$ of synchronous (outgoing) links per new publication    according to a second exponential trend $r(t)=r(0)\exp[g_{r}t]$. For both $n(t)$ and $r(t)$ we use their integer part, and plot their growth in  {\bf Fig. \ref{Figure3.fig}}(a).
We set the initial condition  $r(0) \equiv 25$ in scenarios  featuring no reference list growth (characterized by $g_{r}=0$), such that each new publication  cites 25 prior articles independent of $t$.  Alternatively, in scenarios that do feature reference list  CI, we use the empirical growth rate value, $g_{r}\equiv0.018$  and $r(0)\equiv 5$.
We then sequentially add cohorts of $n(t)$ publications to the  network over $t=1...T \equiv 150$ periods  according to the following link-attachment (citation) rules that capture the salient features of scholarly citation practice:\\

\noindent{ \bf Network growth rules}
\begin{enumerate}
\item {\bf System Growth:} In each period $t$, we introduce $n(t)$ new publications, each citing  $r(t)$ other publications by way of a directed link. Hence, the total number of synchronous (backwards) citations produced in period $t$ is $C(t)=n(t)r(t)$, which grows exponentially at the rate $g_{C} = g_{n} + g_{r}$.
\item {\bf Link Dynamics:}  illustrated in the schematic  {\bf Fig. \ref{Figure3.fig}}(b).  For each new publication $a \in n(t)$:

\subitem {\bf (i) Direct citation $a \rightarrow b$:} Each new publication $a$ starts by referencing  1 publication $b$ from  period $t_{b}\leq t_{a}$ (where $t_{a}=t$ by definition). 
The publication $b$ is selected proportional to its attractiveness, prescribed  by the weight $\mathcal{P}_{b,t} \equiv (c_{\times}+c_{b,t})[n(t_{b})]^{\alpha}$. 
The  factor $c_{b,t}$ is the total number of citations received by $b$ thru the end of period $t-1$,  thereby representing   preferential attachment (PA) link dynamics  \cite{Simon_class_1955,Barabasi_evolution_2002,Jeong_Measuring_2003, Redner2005PA,peterson_nonuniversal_2010}. 
The factor $n(t_{b})$ is the number of new publications introduced in cohort $t_{b}$, and represents   crowding out of old literature by new literature, net of the citation network. The parameter  $c_{\times} \equiv 6$ is a citation offset controlling for the citation threshold, above which preferential attachment ``turns on'' \cite{petersen_reputation_2014} such that a node becomes incrementally more attractive once   $c_{b} \geq c_{\times}$. 

\subitem {\bf (ii) Redirection  $a \rightarrow \{r\}_{b}$:} Immediately after step (i), the new publication  $a$ then cites a random number $x$ of  the publications cited in the  references list  $\{r\}_{b}$ (of size $r_{b}$) of publication $b$.  By definition, $\beta$ represents the fraction of  citations following from this redirection mechanism, which is responsible for the rate of non-spurious triadic closure in the network. Hence,  by construction $\beta = \lambda / (\lambda + 1) \in [0,1]$, where $\lambda$ represents the  average number of citations to elements of $\{r\}_{b}$ by publication $a$ (such that the expected value of $x$ is $\lambda$).
Consequently, $\lambda =\beta/(1-\beta)$ is the ratio of the rate of citations following citing mechanism (ii) by the rate of citation following the `direct' citation mechanism (i). 

We operationalize the stochastic probability of selecting $x$ references according to the binomial distribution, 
\begin{eqnarray}
P(x=k) = {r_{b} \choose k} (q)^{k}(1-q)^{r_{j}-k} \ ,
\end{eqnarray}
 with success rate $q=\lambda/r_{b}$ to ensure that   $\langle x \rangle =\lambda$. 
 Put another way,  on average, the total number of new citations per period  that follow from the redirection citation mechanism (ii) is $r_{(ii)}(t)=\beta r(t)$.

 Once $x$ is determined by way of a random number generator, we then select  $x_{\text{Binomial}(r_{b},q)}$ members from the set  $\{r\}_{b}$ (i.e. without replacement).
Each publication belonging to $\{r\}_{b}$ is selected according to the same weights $\mathcal{P}_{p,t}$ used in step (i). 
As such, this second-stage PA also prioritizes more recent elements of  $\{r\}_{b}$ (i.e., those items with larger $t_{p}$), in addition to more highly-cited elements of $\{r\}_{b}$.
 Note that we do not allow $a$ to cite any given element of $\{r\}_{b}$ more than once within its reference list.
 
\subitem {\bf (c) Stop citing after reaching $r(t)$:} The referencing process alternates between mechanisms (i) and (ii)   until publication $a$ has cited exactly $r(t)$ publications.
\item Repeat step {\it 2. Link Dynamics} for each new publication entering in period $t$. 
\item Update the  PA weights, $\mathcal{P}_{p,t}$, for all existing nodes  at the end of each $t$.
\item Perform steps (1-4) for $t=1...T$ periods and then exit the network growth algorithm. 
\end{enumerate}

\vspace{-0.25in}
\subsection{Computational simulation results}
\vspace{-12pt}
In this section we present the results of a  generative citation network model \cite{pan2016memory}  that incorporates latent features of secular growth and  two complementary citation mechanisms  illustrated in  {\bf Fig. \ref{Figure3.fig}}(b), namely: (i) direct citation from a new publication $a$ to publication $b$; and (ii) redirected citations from $a$ to a random number of publications from the reference list of $b$. 
The redirection mechanism (ii) gives rise to triadic closure in the network, thereby capturing shifts in correlated citation practice -- such as the increased ease at which scholars can follow a citation trail with the advent of web-based hyperlinks, as well as self-citation. This redirection is the dominant contributor to 'consolidation' measured by $N_{j}$ in $CD_{p}$. We explicitly control the rate of (ii) with a tunable parameter $\beta \in [0,1]$ that determines the fraction of links in the citation network  resulting from mechanism (ii). 
And to simulate the net effect of $\beta$, we construct some networks featuring a constant $\beta(t) =0$ and other networks featuring an increasing $\beta(t) \equiv t/400$ such that $\beta (t=150) = 0.375$ corresponding to roughly 1/3 of links arising from mechanism (ii) by the end of the simulation  -- see  {\bf Fig. \ref{Figure3.fig}}(c). 
 
 We construct ensembles of  synthetic networks  according to  six  growth scenarios that incrementally add or terminate either of two citation mechanisms:  $g_{r}=0$ corresponds to no CI; and $\beta = 0$ corresponds to no triadic closure (i.e., no `consolidation'). 
More specifically, the parameters distinguishing the  six  scenarios analyzed in what follows are:
 \begin{enumerate}
  \item[(1)] no CI ($g_{r}=0$ with $r(t)=25$); and no explicit redirection mechanism that controls triadic closure ($\beta = 0$);
  \item[(2)] no CI ($g_{r}=0$ with $r(t)=25$); and an increasing  redirection rate, $\beta(t) = t/400$ such that  $\beta(150) =0.375$;
   \item[(3)] CI implemented using the empirical value ($g_{r}=0.018$) with $r(0)=5$;  and  increasing redirection rate, $\beta(t) = t/400$;
      \item[(4)]  same as (3) but calculated using a larger citation window. 
      \item[(5)]  same as (3) but reference list capped at $r(t) = 25$ for $t\geq T^{*}\equiv 92$.
      \item[(6)]  same as (4) but reference list capped at $r(t) = 25$ for $t\geq T^{*}$.
\end{enumerate}
For each  scenario we  constructed  four distinct synthetic citation networks, each evolving over $t \in [1,T = 150]$ periods (i.e years) from a common initial condition at $t=0$. 
For scenarios (1-3) we calculate $CD_{p}$ using a citation window of CW= 5 periods, whereas in (4) we use CW=10 periods. 
Scenarios (3) and (4)  are shown in order to show the non-linear sensitivity of $CD_{CW}$  to the  CW parameter \cite{bornmann2019disruption}, 
 and demonstrates  that fixed CWs do not  address CI \cite{petersen_citationinflation_2018}.

\begin{figure*}
\centering{\includegraphics[width=0.99\textwidth]{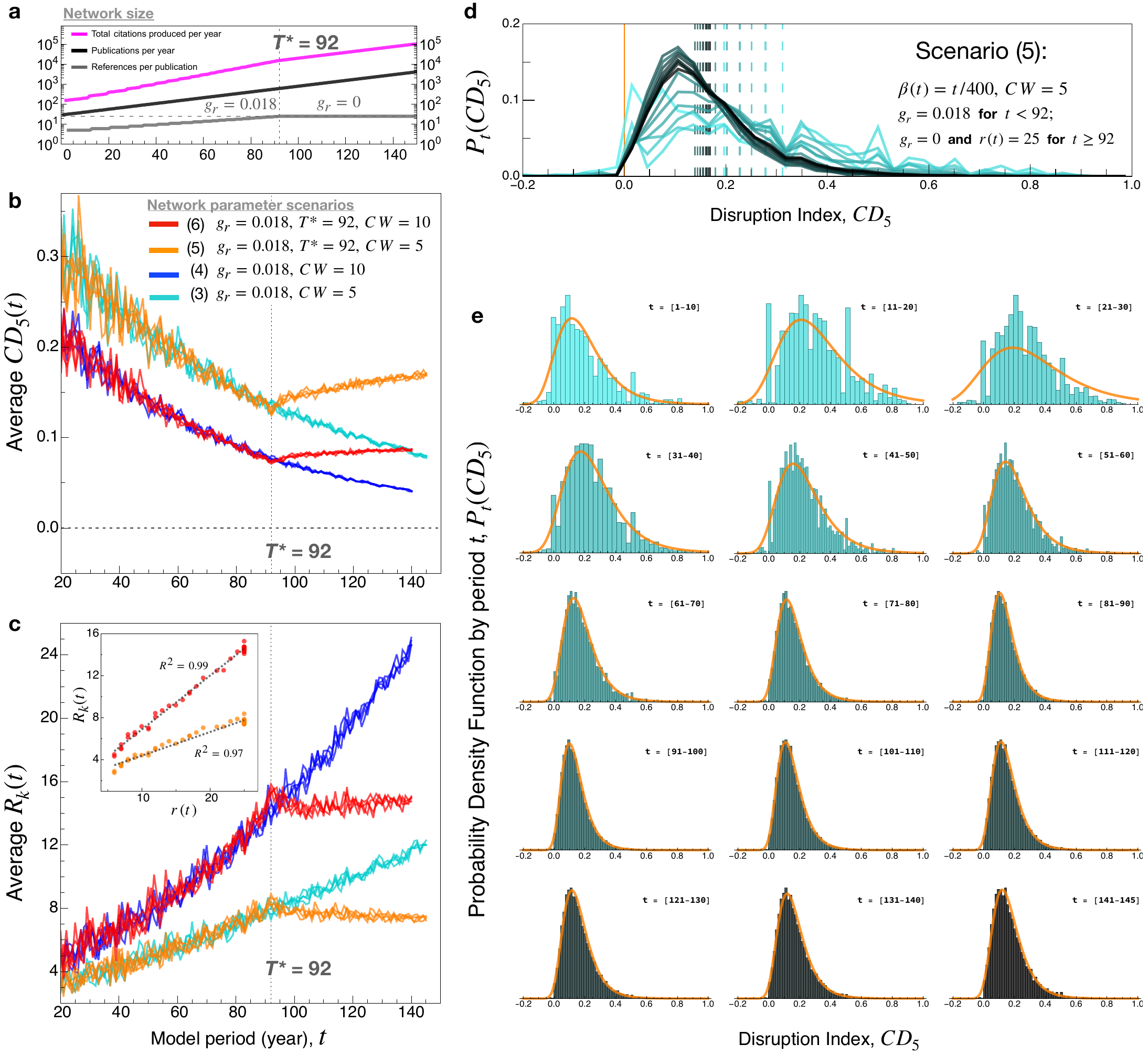}}
 \caption{  \label{Figure5.fig} {\bf Hypothetical publishing policy intervention  reveals effect of capped reference list lengths on $CD$.}
  (a) Evolution of network size in  scenarios (5,6) where the number of references per paper is capped at $r(t\geq T^{*}) = 25$ after $T^{*}=92$, such that the growth in the  total citations produced per year depends solely on the growth of $n(t)$. 
  (b) Average $CD_{CW}(t)$ for scenarios (3)-(6). Immediately after $T^{*}=92$ the $CD(t)$ trends for intervention scenarios (5,6)   reverse from decreasing to increasing. 
  (c) The divergence in $CD(t)$ trends is attributable to the taming of CI which stabilizes $R_{k}(t)$.
 (d)  The frequency distribution $P_{t}(CD_{5})$ aggregated over 10-period intervals indicated by the color gradient; vertical dashed lines indicate distribution mean. 
 Comparing with {\bf Fig. \ref{FigureS2.fig}(a)}, it is clear that the $P_{t}(CD_{5})$ distribution for Scenario (5) becomes significantly more stable  $t \geq T^{*}$, with  variation due to the residual citation inflation associated with publication volume growth ($g_{n}$).   
 (e) The stability of the $P_{t}(CD_{5})$ distribution after $T^{*}$ suggests that quantitative properties of the Extreme Value (Fisher-Tippett) distribution could be used to develop time-invariant disruption measures; orange curves represent the best-fit Fisher-Tippett distribution model.
 }
\end{figure*}

  {\bf Figure \ref{Figure3.fig}}(d)  shows  16 average $CD(t)$ curves calculated for each synthetic network. 
  Because the  sources of network variation are strictly limited to the stochastic link dynamics, there is relatively little variance across each ensemble of networks constructed using  the same scenario parameters, and so in what follows we show all  realizations simultaneously.
  As there are no latent institution, author or other innovation  covariates, then the difference between network ensembles is attributable to either CI or the redirection mechanism.
  
We start by considering  scenarios (1,2) for which $g_{r}=0$, which  show that $CD_{5}(t)$  systematically {\it increases} in the absence of reference list CI. While scenario (1) does capture CI attributable to increased publication volume ($g_{n}>0$),  it does not  appear to be sufficient to  induce a negative trend in $CD_{5}(t)$. 
Scenario (2) features an increasing $\beta(t)$, which results in  larger $CD$ values because redirected citations tend to fall outside shorter CW and thus are not incorporated into the $CD$ subgraph.
Summarily, comparison of (1) and (2)  indicate that the redirection mechanism capturing shifting patterns of scholarly citation behavior is the weaker of the two mechanisms we analyzed.

The comparison of scenarios (2,3) illustrates the role of CI. Notably, scenario (3)  reproduces both the magnitude and rate of the decreasing trend in $CD(t)$   observed for real citation networks  \cite{park2023papers}. 
 {\bf Figure \ref{Figure3.fig}}(e) shows that alternative metric $CD^{\text{nok}}_{5}$ proposed in ref. \cite{bornmann2020disruption}, which also matches the empirical trends reported in ref.  \cite{park2023papers}.
These results demonstrate the acute effect of reference-list CI on $CD$  since the only difference between  scenarios (2) and (3)  pertains to $g_{r}$. 
 
  {\bf Figure \ref{Figure3.fig}}(f)  reproduces  the linear relationship between $r(t)$ and  $R_{k}(t)$ and confirms the empirical relationship shown in {\bf Fig. \ref{Figure2.fig}}(e)  -- thereby solving the mystery regarding the origins of the decreasing disruptiveness over time \cite{kozlov2023disruptive}: as the size of the reference list $\{r\}_{p}$ increases, so does the likelihood that $\{r\}_{p}$ contains a highly-cited paper, which increases $N_k$ to such a  degree that  $R_{k,p} \gg 1$ and so $CD_{p} \rightarrow 0$ independent of the relative differences between disruption and consolidation captured by $N_{i}-N_{j}$. 
  {\bf Figure \ref{Figure3.fig}}(g) shows that even  $CD^{\text{nok}}_{5}$ suffers from  systematic  bias affecting its denominator, and so neglecting the term $N_{k}$ does not solve the fundamental issue of CI.

Scenarios (3,4) reveal the effect of CW, which controls  the size of the set $\{c\}_{p}$ and thus the magnitude and growth rate of $R_{k}(t)$. 
Notably, the number of items included in $\{c\}_{p}$ depends on both CW  and $t$ because the reference age between the cited and citing article increases with time \cite{pan2016memory}. 
 Regardless, the average $CD_{CW}(t) \rightarrow 0$ as $r(t)$ increases, independent of the $CW$ used. 

{\bf Figure  \ref{Figure5.fig}} further explores the implications of CI on $CD$ by modeling a hypothetical scenario in which CI is suddenly `turned off' after a particular intervention time period $T^{*}$.
In this way, scenarios (5,6) explore the implications of a  restrictive publishing policy whereby all journals suddenly agree to impose caps on reference list lengths. Scenarios (5,6) enforce this hypothetical policy  at $t\geq T^{*}\equiv 92$ by way of a piecewise smooth $r(t)$ curve  such that:   $r(t)=r(0)\exp[g_{r}t]$ for $t<T^{*}$ and $r(t) = r(T^{*})=25$ -- see  {\bf Fig. \ref{Figure5.fig}}(a).
 This hypothetical intervention  exhibits the potential for the scientific community to temper the effects of CI by way of strategic publishing policy. 
For completeness,   scenarios (3) and (5) use CW=5 and scenarios (4) and (6) use CW =10. 

{\bf Figures \ref{Figure5.fig}}(b,c) show that  the average $CD(t)$  and $R_{k}(t)$ trajectories for each pair of scenarios are indistinguishable prior to $T*$. Yet immediately after $T*$ the scenarios (5) and (6) diverge from (3) and (4), respectively. 
Notably, the average  $CD_{5}(t)$ in scenarios (5) and (6)   reverse to the point of slowly increasing,  thereby matching  the trends observed for scenario (2).  
In the spirit of completeness, {\bf Fig. \ref{Figure5.fig}}(c) confirms that this trend-reversal is   due to the relationship between $r(t)$ and $R_{k}(t)$.  The shifts in the average $CD_{5}(t)$ are indeed representative of the entire distribution of $CD_{p,5}$ values -- see  {\bf Figs. \ref{Figure5.fig}}(d,e). Interestingly, the distribution $P_{t}(CD_{5})$ converges to  a stable Extreme-Value (Fisher-Tippett) distribution in the absence of reference list growth, which exposes candidate avenues for developing time-invariant measures of disruption by rescaling values according to the location and scale parameters. The feasibility of this approach was previously demonstrated in an effort to develop field-normalized  \cite{UnivCite} and  time-invariant (z-score) citation metrics \cite{petersen2014inequality,HBP_2020}.

\vspace{-0.2in}
\section{Discussion}
\vspace{-0.2in}
In summary, Despite the reasonable logic behind the definition of $CD$, the difference between disruptive and consolidating links appearing in the numerator, ${N_{i}-N_{j}}$, is systematically overwhelmed by the extensive quantity $R_{k} \sim r(t)$ appearing in the denominator of $CD$. 
More specifically, we show that the $CD$ index artificially decreases over time due to citation inflation deriving from ever-increasing $r(t)$, rendering $CD$ systematically biased and unsuitable for cross-temporal analysis. 
For the same reasons that central banks  must design monetary policy to avoid the ill effects of    printing excess money  \cite{orphanides1990money,orphanides2003quest},  researchers analyzing scientific trends  should be wary of  citation-network bibliometrics that are not stable with respect to time.
Scenarios where achievement metrics are non-stationary and thus systematically biased by nominal inflation are common, including researcher evaluation \cite{petersen2010methods}, journal impact factors \cite{ASI:ASI20936},  and even achievement metrics in  professional sports \cite{petersen2011methods,petersen_renormalizing_2020}. 

In addition to the measurement error induced by CI, the disruption index also does not account for confounding shifts in  scholarly citation practice. 
The counterbalance to disruption, captured by the term $N_{i}$ in Eq. (1), is consolidation ($N_{j}$), which  is fundamentally a measure of triadic closure in the subgraph $G_{p}$.  While  triangles may spuriously occur in a random  network, their frequency in real networks is well in excess of random base rates due to the correlated phenomena underlying the scholastic practice -- in particular, increasingly strategic (personal and social) character of scholarly citing behavior. 

The source and implications of citation inflation are not inherently undesirable, and if anything point to  thriving industry emerging from the scientific endeavor. 
The advent of online-only journals is a main reason for the steady increase in $r(t)$, as they are not limited by volume print  capacity, unlike more traditional print journals. Hence, in the  era of  megajournals \cite{petersen2018mismanagement} there may have emerged a tendency to cite more liberally than in the past. 
Another mechanism connecting CI and citation behavior derives from the academic profession becoming increasingly dominated by quantitative evaluation, which thereby  promotes the inclusion of strategic references  dispersed among  the core set of  references directly supporting the research background and findings \cite{abramo2021effects}. Notably, scholars have identified various classes of self-citation \cite{ioannidis2015generalized}, which generally emerge in order to  benefit either the authors \cite{fowler2007does,ioannidis2019standardized,pinheiro2022women}, institutional collectives \citep{tang2015there},   the handling editor \cite{petersen2018mismanagement},  and/or the journal \cite{martin2016editors,ioannidis2019user} -- but are otherwise  difficult to differentiate from `normal' citations.  Regardless of their intent, these self-citations are more likely to contribute to triadic closure because if article $b$ cites $c$ as a result of self-citation, then for the same reason a new article $a$ that cites $b$ (or $c$) is that much more likely to complete the triangle on principal alone.

These two issues -- citation inflation and shifting scholarly behavior --  introduce  systematic  bias in citation-based research evaluation that extends over significant periods of time. Indeed, time is a fundamental confounder, and so to address this  statistical challenge    various methods introducing time-invariant citation metrics  have been developed  \cite{UnivCite,petersen_reputation_2014,petersen_citationinflation_2018,HBP_2020}. 
A broader issue occurs when different variables simultaneously shift over time,  such as the number of coauthors, topical breadth and  depth of individual articles, which makes establishing causal channels between any two variables ever more challenging. By way of example, we analyzed the relationship between $CD_{p}$ and $k_{p}$, using a regression model with fixed effects for publication year to superficially control for secular growth, and observe a positive relationship between these two quantities, in stark contrast to the negative relationship reported in ref. \cite{wu2019large}. 
 
We conclude with a  policy insight emerging from our analysis regarding interventional approaches to addressing  citation inflation. Namely, journals might consider capping reference lists commensurate with the different types of articles they publish, e.g. letters, articles, reviews, etc. An alternative that is more flexible  would be to impose a soft cap based upon the  average number of references per article page \cite{abt2002relationship}.  Results of our computational simulations indicate that such  policy could readily temper the effects of citation inflation in research evaluation, and might simultaneously address other shortcomings associated with self-citations by effectively increasing their cost. 

 \vspace{-0.2in}
\section*{Data Availability}
 \vspace{-0.2in}
 \noindent All synthetic  citation networks analyzed are openly available at the Dryad data repository \cite{DryadDisruption2023}. The psuedocode for the citation network growth is sufficient to generate additional citation networks with different parameters.
 
\bibliographystyle{naturemag}
\bibliography{Bibtex}

\newpage
\clearpage

\beginsupplement

\vspace{-0.2in}
\section{Appendix: Reproduction of  statistical regularities in a real-world citation network -- the Web of Science}
\begin{figure*}
\centering{\includegraphics[width=0.99\textwidth]{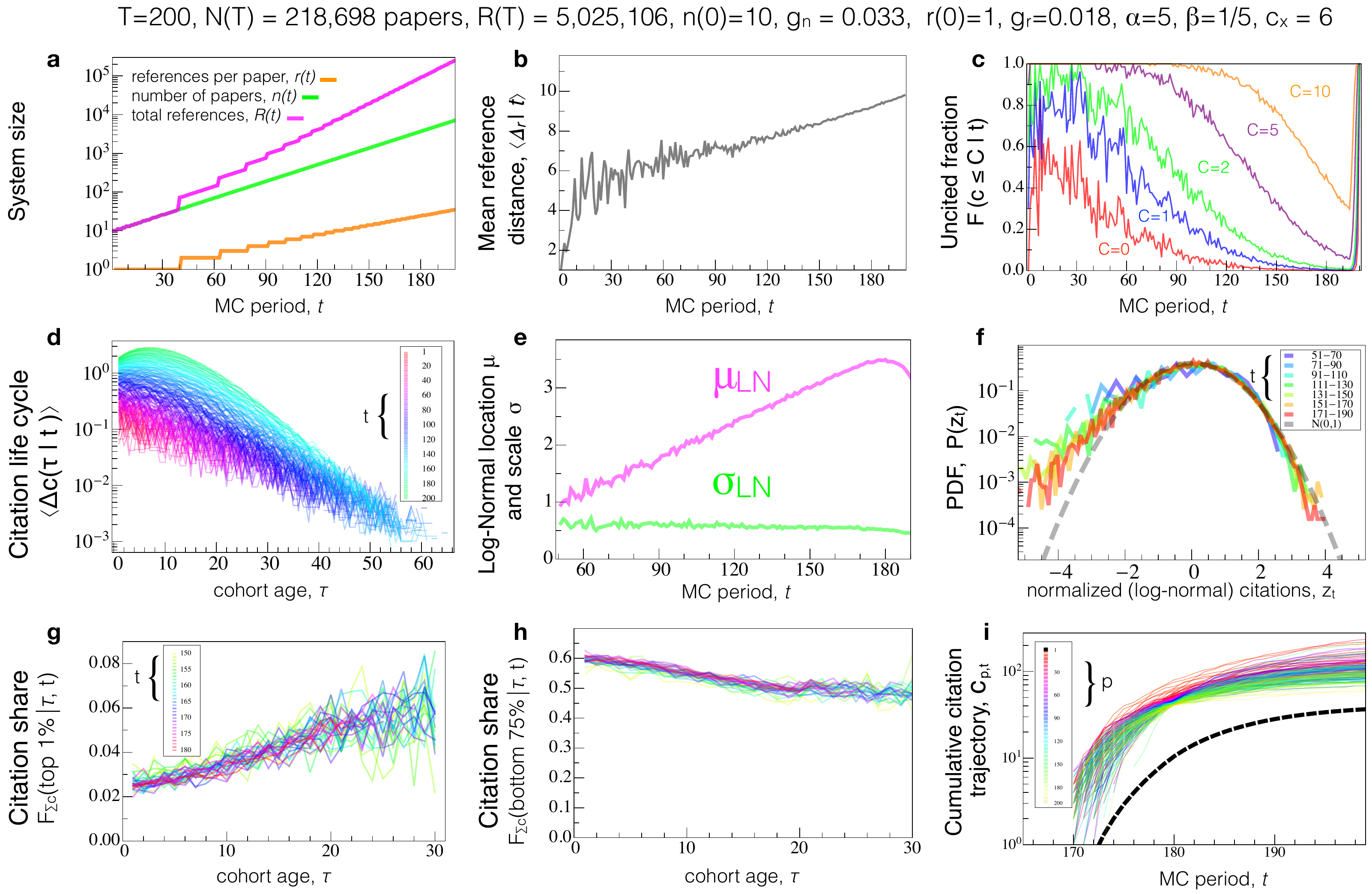}}
\caption{\label{FigureS1.fig}  {\bf Citation redirection model -- reproduction of empirical statistical regularities that characterize real citation patterns.}  
Figure and caption reproduced with permission from  Pan et al. \cite{pan2016memory}. 
Shown are various  properties of the synthetic citation  network that can be compared with empirical trends. We evolved the simulation using the parameters: $T\equiv$  200  MC periods ($\sim$ years), $n(0)\equiv$  10 initial publications, $r(0)\equiv$ 1 initial references, exponential growth rates $g_{n} \equiv 0.033$ and $g_{r} \equiv 0.018$, secondary redirection parameter $\beta  \equiv 1/5$ (corresponding to $\lambda=1/4$), citation offset $C_{\times}\equiv6$, and life-cycle decay factor $\alpha \equiv 5$. 
At the final period $t=T$, the final cohort has size $n(T)=7112$ new publications, $r(T)=35$ references per publication, and final citation network size  
 $N(200)$ = 218,698  publications (nodes) and $R(T)$ = 5,025,106  total references/citations (links). (a) The size of the system in each MC period $t$. (b) Growth of the  mean reference distance $\langle \Delta_{r} \rangle$. 
 (c) The fraction $f_{c\leq C}(t \vert \tau=5)$ of publications which have $C$ or less citations at cohort age $\tau=5$.
 (d) The citation life cycle, measured here by the mean number of new citations  $\tau$ periods after entry (publication). The different curves correspond to the publication cohort entry period $t$. For sufficiently large $t$ the life cycle decays exponentially. 
 (e) Growth of the logarithmic mean (location) value  $\mu_{LN,t}$ and the relative stability of the logarithmic standard deviation (scale) value $\sigma_{LN,t}$.   $\mu_{LN,t}=\langle \log (c_{p,t}) \rangle$ and $\sigma_{LN,t}= \sigma[ \log (c_{p,t})]$ are the logarithmic mean and standard deviation calculated across all $p$ within each age cohort $t$.
 (f) The distribution $P(z_{p,t})$ of the normalized citation impact $z_{p,t}$. For visual comparison  we plot the Normal distribution $N(\mu=0,\sigma=1)$.  (G) The increasing  citation share $f_{\sum c}$ -- the fraction of the total citations received by all publications from cohort $t$  -- of the top $1\%$ of publications  from cohort $t$ (ranked at cohort age $\tau=10$). 
 (h) The decreasing citation share $f_{\sum c}$  of the bottom $75\%$ of publications. 
  (i) The cumulative citation count $c_{p}(t)$ of the top 200  publications $(p)$ from the interval $t=[170,179]$, ranked according to $c_{p}(t=180)$. The dashed line represents the average citations for $p$ from the same cohort over the same period.}
\end{figure*}

\begin{figure}
\centering{\includegraphics[width=0.69\textwidth]{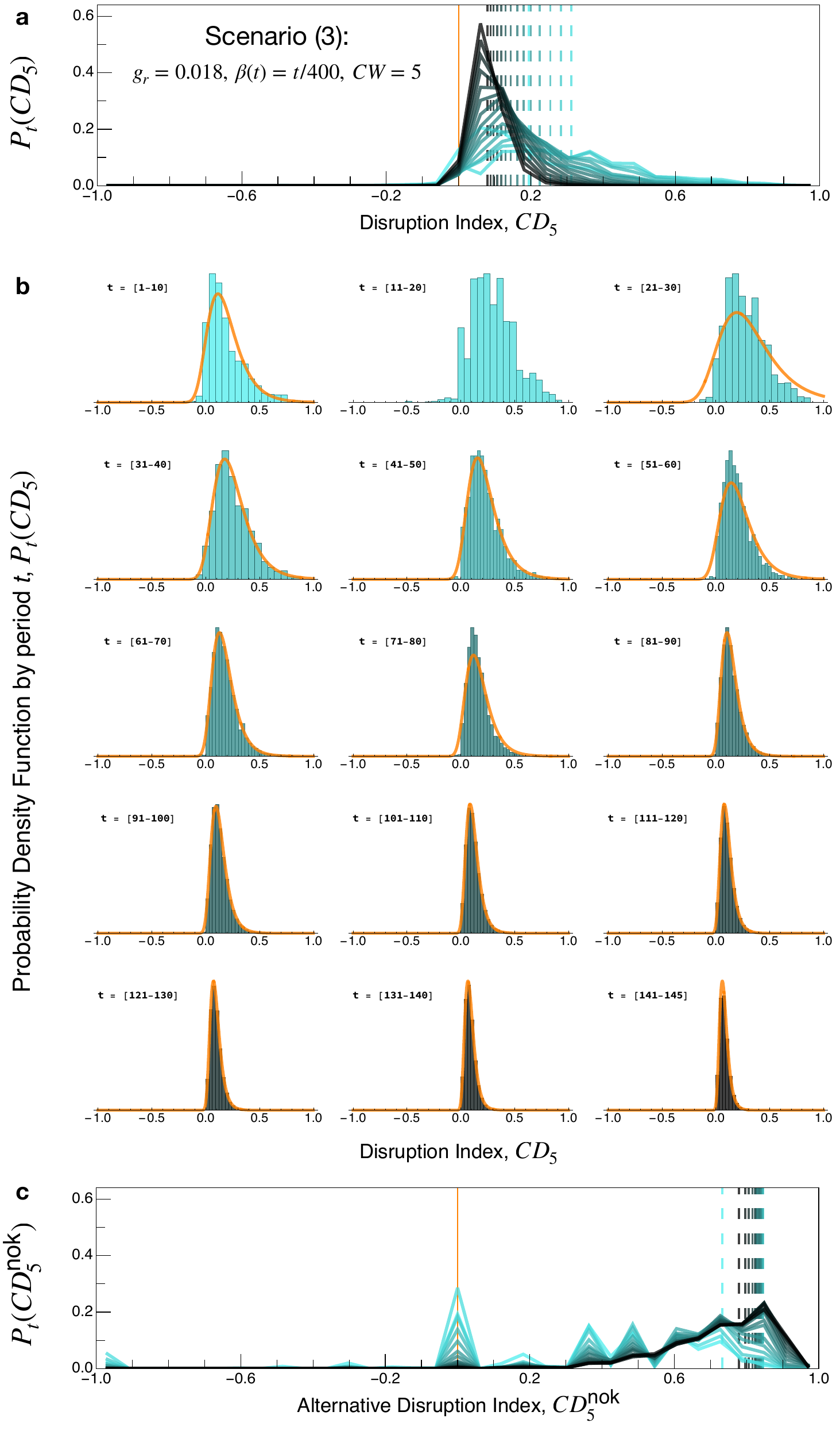}}
 \caption{  \label{FigureS2.fig} {\bf The  distribution of $CD_{5}$ derived from synthetic citation networks  follows the Extreme Value (Fisher-Tippett) distribution but is not stable over time.}
 (a) The probability density function $P_{t}(CD_{5})$ is calculated using values aggregated over 10-period intervals indicated by $t$, with color gradient  indicating each 10-period interval. Vertical dashed lines indicate distribution mean. 
 (b) Each 10-period $P_{t}(CD_{5})$ is shown  with the best-fit Extreme Value (Fisher-Tippett) distribution (orange curve), estimated using Mathematica 13.1 algorithm {\it FindDistribution}. The Extreme Value distribution is a better fit as $t$ increases, pointing to a strategy for normalizing $CD_{5}$ that supports cross-temporal analysis in the same way that the properties of the log-normal distribution can be used to  normalize citation counts collected over different periods \cite{petersen_reputation_2014,petersen_citationinflation_2018,HBP_2020}. (c) Distribution of an alternative disruption index, $P_{t}(CD^{\text{nok}}_{5} )$, calculated using same temporal periods  as in (a), shows that the vast majority of publications according to this measure are highly disruptive.
}
\end{figure}

\vspace{-0.2in}
The following is a summary of the structural and dynamical regularities that characterize a  typical  network produced by our model using the growth parameters indicated along the top bar of {\bf Fig. \ref{FigureS1.fig}}. In addition to the stylized regularities listed below, the citation model also reproduces the temporal trends in $CD_{5}$, $CD_{5}^{\text{nok}}$, and the frequency distribution $P(CD_{5})$, reported by Park et al. \cite{park2023papers} -- see  {\bf Figs \ref{Figure1.fig}}, {\bf \ref{FigureS1.fig}} and {\bf \ref{FigureS2.fig}}.

{\bf Figure \ref{FigureS1.fig}(a)}  shows the time series  $n(t)$, $r(t)$, and $R(t)$  as determined by the empirical parameters $g_{n}$, $g_{r}$, and $g_{R}$. 

{\bf Figure \ref{FigureS1.fig}(b)} shows the mean of the reference distance $\Delta_{r} = t_{a} - t_{p}$ calculated as the time difference between the publication year of $a$ and of any given publication $p$ that it cites. The increasing $\langle \Delta_{r} \vert t \rangle$ conforms with prior theoretical and empirical work \cite{egghe2010model,lariviere_long-term_2008,acharya_rise_2014,pan2016memory}.
 
 {\bf Figure \ref{FigureS1.fig}(c)} shows the decreasing frequency of publications with less than $C$ = 0, 1, 2,5, 10 citations. This trend is consistent with empirical work \cite{schwartz_rise_1997,Wallace2009296,lariviere_decline_2009}, and  has profound implications on the connectivity of the citation network, and search and retrieval algorithms based upon the connectivity.
 
 {\bf Figure \ref{FigureS1.fig}(d)} shows the average citation life-cycle, $\Delta c (\tau  \vert t) $ of individual publications conditioned on their publication year $t$, where $\tau$ is the age of the publication in that year, $\tau_{p} = t-t_{p}+1$. The exponential decay of the consistent with empirical work \cite{parolo_attention_2015}.
  
 {\bf Figure \ref{FigureS1.fig}(e)} shows the mean and standard deviation of $c' = \ln (c_{p}+1)$, where citations counts  $c_{p}$ are tallied at $T$, the final period of the model. 
Naturally, very recent publications have not had sufficient time to accrue citations. Also, very early publications were at the peak of their lifecycle during periods in which there was smaller $n(t)$. Hence, the average $\mu_{LN}$ peaks near the end of the model, and then decays to 0 for the final period. This systematic bias due to citation inflation, as well as the right-censoring bias, may seem difficult to overcome. However, the location and scale given by $\mu_{LN}$ and $\sigma_{LN}$, respectively, provide a powerful solution, which is to normalized citations according to the rescaling,
\begin{eqnarray}
z_{p,t}= \frac{\ln(c_{p,t}+1)-\mu_{LN,t}}{\sigma_{LN,t}} \ ,
\end{eqnarray}
where  $\mu_{LN,t}=\langle \ln (c_{p,t}+1) \rangle$ and $\sigma_{LN,t}= \sigma[ \ln (c_{p,t}+1)]$ are the  mean and the standard deviation of the logarithm of $c_{p,t}+1$ calculated across all $p$ within each  $t$.
 This normalization procedure leverages the property that the distribution $P(c_{t}\vert t)$ is log-normally distributed, as shown for real citation networks \cite{UnivCite}. As such, the distribution $P(z_{t})$ takes the form of a standardized z-score distributed according to the normal distribution $N(0,1)$, which  is stable over time. As shown in  {\bf Fig. \ref{FigureS1.fig}(f)}, $P(z)$ forms an inverted parabola when plotted on log-linear axes, independent of $t$. This normalization is useful in regression settings aimed at identifying citation effects net of temporal trends, where $t$ is included in the model specification as either as a continuous or dummy variable  \cite{petersen_citationinflation_2018,HBP_2020}. 
 
   {\bf Figures \ref{FigureS1.fig}(g,h)} show   the evolution of the citation share of the top and bottom percentile groups $F_{\sum c}(Q | \tau, t)$, consistent with empirical work  showing that a small fraction of the top-cited papers from high-impact journals increasingly dominate the future citations of that journal \cite{Barabasi2012}.
   
 And {\bf Figure \ref{FigureS1.fig}(i)} shows individual citation trajectories, $c_{p,t}$, produced by the model. The shape and distribution of the cohort are consistent with empirical citation trajectories reported in \cite{petersen_reputation_2014}.

\end{document}